\documentclass[aps,floatfix,10pt,amssymb,amsfont,amsmath,superscriptaddress,float,nofootinbib,longbibliography,nobalancelastpage,twocolumn]{revtex4-2}

\usepackage{color}
\usepackage{graphicx}
\usepackage{hyperref}
\usepackage{braket}
\usepackage{tikz}
\usepackage{yquant}
\usepackage[export]{adjustbox}
\hypersetup{hidelinks}
\begin{document}


\author{Oliver O'Brien}
\affiliation{Department of Applied Mathematics and Theoretical Physics, University of Cambridge,\\ Wilberforce Road, Cambridge, CB3 0WA, United Kingdom}
\author{Atsushi Ueda}
\affiliation{Department of Physics and Astronomy, Ghent University, Krijgslaan 281, 9000 Gent, Belgium}
\author{Etienne Granet}
\affiliation{Quantinuum, Leopoldstraße 180, 80804 Munich, Germany}
\author{Henrik Dreyer}
\affiliation{Quantinuum, Leopoldstraße 180, 80804 Munich, Germany}
\author{Frank Verstraete}
\affiliation{Department of Applied Mathematics and Theoretical Physics, University of Cambridge,\\ Wilberforce Road, Cambridge, CB3 0WA, United Kingdom}
\affiliation{Department of Physics and Astronomy, Ghent University, Krijgslaan 281, 9000 Gent, Belgium}

\title{Monopoles, Scattering and Generalized Symmetries on a Quantum Computer}
\begin{abstract}
Quantum computers have advanced to the stage that interesting many-body quantum phenomena can be observed. A striking example is the magnetic monopole paradox, a problem that was only solved recently using the language of generalized symmetries. Here we demonstrate a quantum spin chain analogue of magnetic monopole scattering \cite{ueda2025} on Quantinuum System Model H2. We implement a novel protocol for preparing a localized quasiparticle wavepacket on top of the ground state. By combining this protocol with time evolution across a duality defect, we efficiently convert local particles into non-local string-like excitations. We demonstrate this particle-to-string conversion in the transverse-field Ising model and a symmetry-protected topological (SPT) phase of the cluster model. Our results establish duality defects and generalized symmetries as practical resources for preparing and probing non-local excitations on quantum hardware.
\end{abstract}

\maketitle

\section{Introduction}
Understanding quantum many-body systems is a central goal of physics, but strong correlations often place them beyond the reach of mean-field theory or perturbative methods. Scattering offers a direct way to learn about them: prepare a low-energy excitation above the ground state, let it scatter, and examine what comes out. The outgoing state reveals both the excitations the system supports and how they interact.

Recognizing the outgoing excitation, however, can be surprisingly subtle. At certain interfaces between regions of a system, an excitation created by a local operator may pass through without reflection, yet emerge as the endpoint of a nonlocal string~\cite{ueda2025}. The excitation keeps moving, but following it requires knowing what to measure. Without the right observables, a particle that continues to propagate can appear to ``disappear.''

An apparent disappearance can also arise when the outgoing state falls outside the scattering channels being considered. In familiar scattering problems, the missing probability can often be recovered by including ordinary multiparticle states. A more puzzling disappearance arose in the work of Callan and Rubakov on chiral fermions scattering from magnetic monopoles: the conventional fermionic description appeared to offer no outgoing state consistent with the conserved quantum numbers~\cite{Callan:1982ah,Rubakov:1982fp,Polchinski:fermion_rotor}. In this sense, an incoming fermion seemed unable to scatter back from the monopole. Of course, it is not that quantum mechanics
has failed. The difficulty lies in identifying the outgoing state, a problem that became known as the “monopole paradox.”

Recent work has proposed a resolution in terms of twisted sectors~\cite{van_beest_fermion-monopole_2024,van_beest_monopoles_2025,loladze_monopole-fermion_2025,loladze_dynamics_2025}. In the effective one-dimensional description of radial scattering, the monopole acts as a boundary whose degrees of freedom participate in the scattering. The outgoing excitation belongs to a twisted sector and is attached to the monopole by a topological defect line. A unitary description therefore includes both the monopole degrees of freedom and the relevant twisted sectors.

This structure has a direct lattice analogue in spin chains with topological duality defects~\cite{ueda2025,Ueda:2026wvc,aasen2016topological}. Such a defect joins two dual descriptions of a system, relating their excitation spectra and dynamics. An incoming local excitation can be perfectly transmitted while acquiring a nonlocal representation, with a topological string extending back to the interface. As in the monopole problem, the defect degrees of freedom participate in this conversion and remain part of the full scattering description. These lattice models thus bring this shared mechanism into a setting accessible to quantum hardware~\cite{samanta2025dualitytwist}.

Realizing this conversion experimentally begins before the particle reaches the defect. Even when a many-body ground state is available, preparing a localized wavepacket with a controlled mean momentum requires a coherent superposition of excitations at different sites. The corresponding operator is generally nonunitary, and its implementation can incur post-selection overhead or require feed-forward~\cite{childs2012hamiltonian,turco2025creation,Gily_n_2019,mcardle2026quantum,o2025quantum,corcoles2021exploiting,pino2021demonstration,farrell2026digital,davoudi2024scattering}. Direct preparation of a nonlocal wavepacket adds the task of coherently combining extended string operators. A duality defect lets the dynamics do some of this work: we prepare the excitation where it is local and allow scattering to generate its nonlocal representation. Following the excitation through this change then requires measurements suited to the outgoing state.

Here, we implement this strategy on Quantinuum's H2 quantum computer~\cite{PhysRevX.13.041052}. Our novel ancilla-assisted preparation protocol identifies each wavepacket's momentum sector from the measured ancilla outcome, retaining all valid outcomes without feed-forward. Combining this preparation with local and string observables, we reconstruct momentum-resolved space--time trajectories in two complementary lattice constructions. In the Ising model, we demonstrate that duality defects allow us to switch a propagating excitation back and forth between local and nonlocal descriptions. In the cluster construction, the outgoing excitation slips out of view of the original local spin observable, while its trajectory remains visible in nonlocal string-order measurements. The defect thus serves as both an object of study and a tool for preparing nonlocal excitations, giving duality defects and generalized symmetries a practical role on quantum hardware.

Beyond the duality-defect setting, our novel wavepacket preparation protocol is a useful primitive in its own right. It ensures every shot provides physically meaningful data and it requires no variational optimization of the excited many-body state. This simplicity makes it a natural starting point for future dynamics experiments.

In the following, we introduce the wavepacket preparation protocol in Section~\ref{section:localized_wavepacket} and construct the lattice models with duality defects in Section~\ref{section:nonlocalwavepacket}. The hardware implementation and measured scattering dynamics are presented in Sections~\ref{section:hardwareimplementation} and~\ref{section:results}, respectively.

\section{Momentum-resolved wavepacket preparation}
\label{section:localized_wavepacket}
Preparing a quasiparticle wavepacket excitation on top of a many-body ground state on a quantum computer is challenging. One first needs to prepare a many-body ground state $|\Omega\rangle_S$ and then apply a non-unitary operator to create an excitation. For a Gaussian wavepacket, for instance, this operator is of the form
\begin{equation}
    \Psi_k^{\dagger} = \sum_{x=0}^{M-1} a_xe^{\frac{2\pi i}{M}kx} \phi^{\dagger}_x\,,
\end{equation}
where $\phi^{\dagger}_x$ are unitary creation operators for localised excitations and $a_x$ are gaussian real coefficients 
$$a_x=\frac{1}{\mathcal{N}}e^{-\left(\frac{x-x_0}{\sigma}\right)^2}\,,$$
where $\mathcal{N}$ and $\sigma$ are a normalization constant and the gaussian width, respectively. Then our goal is to apply this operator to the ground state as 
$$|\psi_k\rangle_S = \Psi_k^{\dagger}|\Omega\rangle_S\,.$$
Since $\Psi_k^\dagger$ is not unitary, the states $|\psi_k\rangle_S$ are in general unnormalised, with
\begin{equation}
\| \ket{\psi_k} \|^2 = \sum_{x,y=0}^{M-1} a_x a_y\, e^{\frac{2\pi i}{M}k(x-y)} \braket{\Omega | \phi_y \phi_x^\dagger | \Omega}_S ,
\end{equation}
which reduces to $\| \ket{\psi_k} \|^2 =\sum_x a_x^2 = 1$ only when $\braket{\Omega | \phi_y \phi_x^\dagger | \Omega}_S = \delta_{xy}$ (e.g. when acting on the vacuum). 

In general, implementing such a non-unitary operator on a quantum computer requires post-selection or mid-circuit measurement with feed-forward \cite{childs2012hamiltonian, turco2025creation, Gily_n_2019, mcardle2026quantum, o2025quantum, corcoles2021exploiting, pino2021demonstration, turco2025creation}. These approaches incur additional cost in either shot overhead or architectural complexity. We instead use a post-classification technique that retains every valid shot by preparing the state
\begin{equation}
    \frac{1}{\sqrt{M}}\sum_{k=0}^{M-1} \ket{k}_A\otimes|\psi_{k_0+k}\rangle_S\,.
\end{equation}
Measuring the ancilla register to be $k$ conditions the system on the application of $\Psi^{\dagger}_{k+k_0}$. We therefore prepare momentum-resolved wavepackets without discarding valid shots and without feed-forward. Although the momentum is not selected in advance, each shot can be assigned to a momentum sector from the ancilla outcome.

\begin{figure}[tb]
    \centering
\includegraphics[width=\linewidth]{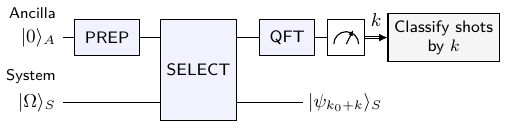}
    \caption{Momentum-resolved wavepacket preparation by ancilla post-classification. The measured ancilla outcome $k$ labels the conditional system state $|\psi_{k_0+k}\rangle_S$. All valid outcomes are retained without feed-forward.}
    \label{fig:qcircuit}
\end{figure}

The circuit (shown in Fig.~\ref{fig:qcircuit}) is closely related to linear combinations of unitaries \cite{childs2012hamiltonian, berry2015simulating, babbush2018encoding} and consists of the following steps:
\begin{enumerate}
    \item \textsc{PREP}: On $m = \log_2 M$ ancillas initialised to $\ket{0}_A$ prepare the following state: $
            \sum_{x=0}^{M-1} a_xe^{\frac{2\pi i}{M}k_0x} \ket{x}_A$
    \item \textsc{SELECT}: Controlled upon this ancilla register act with $\phi^{\dagger}_x$ upon the system giving:
    $\sum_{x=0}^{M-1} a_xe^{\frac{2\pi i}{M}k_0x} \ket{x}_A\otimes \phi^{\dagger}_x\ket{\Omega}_S
    $
    \item \textsc{QFT}: Apply the quantum Fourier transform to the ancilla register, giving the final state
\end{enumerate}
\begin{align*}
       \frac{1}{\sqrt{M}} &\sum_{x,k=0}^{M-1} a_xe^{\frac{2\pi i}{M}(k_0+k)x} \ket{k}_A\otimes \phi^{\dagger}_x\ket{\Omega}_S \\
       &=\frac{1}{\sqrt{M}}\sum_{k=0}^{M-1}  \ket{k}_A\otimes \ket{\psi_{k_0+k}}_S\,.
\end{align*}

Many alternative approaches to preparing localised wavepackets exist. These include (but are not limited to): adiabatically evolving a wavepacket of a free theory into one in an interacting theory \cite{jordan2012quantum, jordan2011quantum, jordan2014quantum, brennen2014multi, barata2021single}, performing symmetry constrained variational optimization on a state with definite momentum \cite{van2021real,farrell2026digital} or straight-forward variational optimization \cite{farrell2024quantum, davoudi2024scattering, zemlevskiy2025scalable}. We do not claim to outperform the state-of-the-art in wavepacket preparation, but believe our methodology may prove a valuable additional tool in particular circumstances. For example, for large systems that prove resistant to variational quantum eigensolver techniques but for which we can efficiently prepare an approximation to the ground state, our approach of applying the wavepacket on top of a prepared ground state may be the most appropriate choice.
\section{Scattering through duality defects}
\label{section:nonlocalwavepacket}
A duality defect changes the operator description of an excitation without changing the underlying propagating degree of freedom. Consequently, a wavepacket created by local operators on one side of the interface may be represented by a string operator on the other. 

Our experiment exploits this property directly. We prepare the incident wavepacket in a region where it is local, evolve it towards the defect, and reconstruct the outgoing excitation using observables appropriate to the dual region. This avoids direct preparation of a superposition of extensive strings.

Generalised symmetries provide the formal structure for this construction~\cite{shao2023noninvertible}. A model with non-local excitations can admit a dual description in which the same excitations are localised \cite{lootens2025entanglement}. In one dimension, the intertwiner between such dual models can be represented as a matrix-product operator (MPO) \cite{lootens2023dualities, lootens2024dualities}. Joining the two descriptions with the corresponding defect yields an interface across which the ideal wavepacket is perfectly transmitted \cite{ueda2025}. The defect can therefore be viewed as a boundary between different topological sectors, analogous to the role played by the monopole in the motivating scattering problem~\cite{khan2024quantumoperations}.

\subsection{Impurity models}
We realise the scattering process in two complementary models: The Ising and cluster models. 

In the Ising example, we construct a chain with three regions: a paramagnetic (PM) region of the transverse-field Ising model, its KW dual in the ferromagnetic (FM) phase, and a second PM region. The adjacent regions are connected by KW duality defects, giving the schematic structure
$$\text{Ising:}\,\mathrm{PM} \;\xrightarrow{\mathrm{KW}}\; \mathrm{KW-dual\ Ising} \;\xrightarrow{\mathrm{KW}}\; \mathrm{PM} $$
A wavepacket created by local operators in the first PM region becomes a domain-wall wavepacket upon crossing the first defect. Whereas direct preparation of this domain-wall wavepacket would involve nonlocal string operators, here it is generated dynamically by scattering through the defect. Crossing the second defect returns the excitation to its original local representation, completing a particle-to-domain-wall-to-particle conversion.

In the cluster-model construction, we begin with two interleaved Ising chains, each with a PM region connected to its KW-dual FM region. In the third region, we apply the Kennedy–Tasaki (KT) transformation to the two decoupled FM chains, obtaining a cluster-model Hamiltonian whose ground state lies in a symmetry-protected topological (SPT) phase~\cite{else2013hidden}. The resulting structure is
$$\text{Cluster:}\,2\times\mathrm{PM} \;\xrightarrow{\,2\times\mathrm{KW}\,}\; 2\times\mathrm{KW\!-\!dual\ Ising} \;\xrightarrow{\mathrm{KT}}\; \mathrm{SPT}.$$
A wavepacket created locally in the PM region first becomes a domain-wall wavepacket in the FM region and then a string excitation in the SPT region. Unlike the Ising construction, the second defect does not restore the excitation’s original local representation. Instead, the propagating excitation becomes {\it{invisible}} to the original local spin observable while remaining detectable through the nonlocal string order parameter.

\subsubsection{Ising Model: conversion into a domain wall}
The explicit Hamiltonian is
$$H_{\mathrm I}=H_{\mathrm p}^{(1)}[\Lambda_{\mathrm{out}}]+H_{\mathrm{dw}}^{(1)}[\Lambda_{\mathrm{mid}}]
+H_{\mathrm{im}}^{(L)}
+H_{\mathrm{im}}^{(R)},$$
where 
\begin{align*}
H_{\mathrm{p}}^{(r)}[\Lambda]
&=
-\sum_{j\in\Lambda} Z_j
-g\sum_{\substack{j\\ j,j+r\in\Lambda}}
X_jX_{j+r},
\\
H_{\mathrm{dw}}^{(r)}[\Lambda]
&=
-\sum_{\substack{j\\ j,j+r\in\Lambda}}
Z_jZ_{j+r}
-g\sum_{j\in\Lambda}X_j,\\
H_{\mathrm{im}}^{(L)}
&= -Z_{I_{L}} Z_{I_{L}+1}
-g X_{I_{L}-1} X_{I_{L}}. \\
H_{\mathrm{im}}^{(R)}
&= -Z_{I_{R}-1} Z_{I_{R}}
-g X_{I_{R}} X_{I_{R}+1}. \\
\end{align*}
$\Lambda_{\mathrm{out}}=[0,I_L-1]\cup[I_R+1,L-1]$ and $\Lambda_{\mathrm{mid}}
=[I_L+1,I_R-1].$ $H_\mathrm{p}^{(r)}$ is an $ r$-decoupled transverse-field Ising model, and $H_\mathrm{dw}^{(r)}$ is its dual model. The overall Hamiltonian $H_\mathrm{I}$ is obtained by applying this duality to a transverse-field Ising model with two maximally entangled defect sites, as illustrated below:

\begin{equation*}
    \centering
    H_I =
    \includegraphics[width=0.35\linewidth, page=2, valign=c]{figs/figs.pdf}
    =
    \includegraphics[width=0.25\linewidth, page=3, valign=c]{figs/figs.pdf}
\end{equation*}

The two regions describe the same underlying excitation in different sectors. In $H_\mathrm{p}$, the incident wavepacket is created by a superposition of local Pauli-$X$ operators. After crossing a KW defect into $H_\mathrm{dw}$, the corresponding excitation is a propagating domain wall and is therefore non-local. The defect thus converts a local particle into a non-local particle while preserving its propagation through the system.

Because the KW duality preserves the spectrum and dynamics of the original model, the ideal construction predicts perfect transmission across both defect sites. The outgoing state is not a different wavepacket created at the boundary, but the dual representation of the incident excitation. The wavepacket subsequently encounters a second KW defect; crossing this interface reverses the transformation and returns the excitation to a local representation. The full trajectory therefore realises a particle-to-string-to-particle scattering sequence.

\subsubsection{Cluster model: conversion into an SPT string excitation}
In the Ising construction, the outgoing domain wall requires a non-local superposition of operators for direct preparation but remains detectable through a local order parameter. We now introduce a three-region model that separates propagation from local visibility more sharply: after crossing the second defect, the excitation propagates on top of the cluster state and is detected through non-local string-order parameters rather than the original local observables:

$$
H_{\mathrm C}=H_{\mathrm p}^{(2)}[\Lambda_L]+H_{\mathrm{dw}}^{(2)}[\Lambda_M]+H_{\mathrm{SPT}}[\Lambda_R]+H_{\mathrm{im}}^{(L)}+H_{\mathrm{im}}^{(R)},
$$
where 
\begin{align*}
H_{\mathrm{SPT}}[\Lambda]
&=
-\sum_{\substack{j\\ j,j+1,j+2\in\Lambda}}
Z_jX_{j+1}Z_{j+2}
-g\sum_{j\in\Lambda}X_j.\\
H_{\mathrm{im}}^{(L)}
&=
- Z_{I_L}Z_{I_L+2}
- Z_{I_L+1}Z_{I_L+3}\\
&\quad- g X_{I_L-2}X_{I_L}
- g X_{I_L-1}X_{I_L+1},
\\
H_{\mathrm{im}}^{(R)}
&=
- Z_{I_R-2}Z_{I_R}Z_{I_R+1}
- Z_{I_R-1}X_{I_R}X_{I_R+1}Z_{I_R+2}
\end{align*}
with $\Lambda_L=[0,I_L-1], \Lambda_M=[I_L+2,I_R-1],\Lambda_R=[I_R+1,L-1].$
This model consists of a pair of interleaved open-boundary transverse-field Ising chains transformed by three dualities. First, two copies of KW duality are applied to the left-hand side of the chains, relating their paramagnetic and ferromagnetic descriptions as in the preceding construction. Second, a spin-$\frac{1}{2}$ analogue of the Kennedy-Tasaki transformation \cite{li2023noninvertible} is applied to the right-hand side of both chains, producing a single $ZXZ$ cluster model. The Kennedy-Tasaki transformation is represented by an MPO constructed from two tensors:
\begin{align*}
    \includegraphics[width=0.2\linewidth,page=4, valign=c]{figs/figs.pdf} = \sum_{a,j=0}^1(-1)^{aj} \ket{a}\bra{a} \otimes \ket{j}\bra{j}\\
    \includegraphics[width=0.25\linewidth,page=5, valign=c]{figs/figs.pdf} = \sum_{a,j=0}^1 \ket{a\oplus j}\bra{a} \otimes \ket{j}\bra{j}\,.
\end{align*}
These combine together into the following Kennedy-Tasaki MPO:
\begin{equation*}
    \includegraphics[width=0.4\linewidth,page=6]{figs/figs.pdf}
\end{equation*}
Applying this transformation to the pair of interleaved open-boundary Ising chains, after KW duality has been applied on their left-hand side, produces $H_c$:
\begin{equation*}
    H_{c}=\includegraphics[width=0.6\linewidth, page=8, valign=c]{figs/figs.pdf}
\end{equation*}

The $H_\mathrm{SPT}$ cluster model realises a one-dimensional $\mathbb{Z}_2\times \mathbb{Z}_2$ SPT phase whose ground state at $g=0$ is the cluster state. This SPT phase extends to $g<1$, where the ground state is no longer the simple fixed-point cluster state. A wavepacket created locally in $H_\mathrm{p}$ first becomes a domain wall in $H_\mathrm{dw}$ and then a non-local string excitation propagating through the SPT phase in $H_\mathrm{SPT}$. Unlike the Ising construction, the two defects implement different transformations, so the outgoing excitation does not return to its original local description.

This construction provides a direct experimental test of whether a particle that disappears from local observables has been destroyed or has instead entered a non-local sector. Before reaching the second defect, the wavepacket can be tracked using the original local observables. After entering $H_\mathrm{SPT}$, the same propagating excitation cannot be detected by the original local observables but is detected using the non-local string order inherited from the duality transformation.

\subsection{Experimental implementation}
\label{section:hardwareimplementation}
To implement the scattering dynamics of both models on quantum hardware, we
\begin{enumerate}
    \item prepare an approximation to the many-body ground state,
    \item excite a local wavepacket on top of this background,
    \item evolve the wavepacket through the duality defects, and
    \item measure the outgoing excitation using the appropriate local or non-local observable.
\end{enumerate}
Each step is adapted to reduce the circuit depth and measurement cost on current quantum hardware as discussed in Appendix~\ref{app:experimental_implementation}.

\begin{figure}
    \centering
    \includegraphics[width=\linewidth]{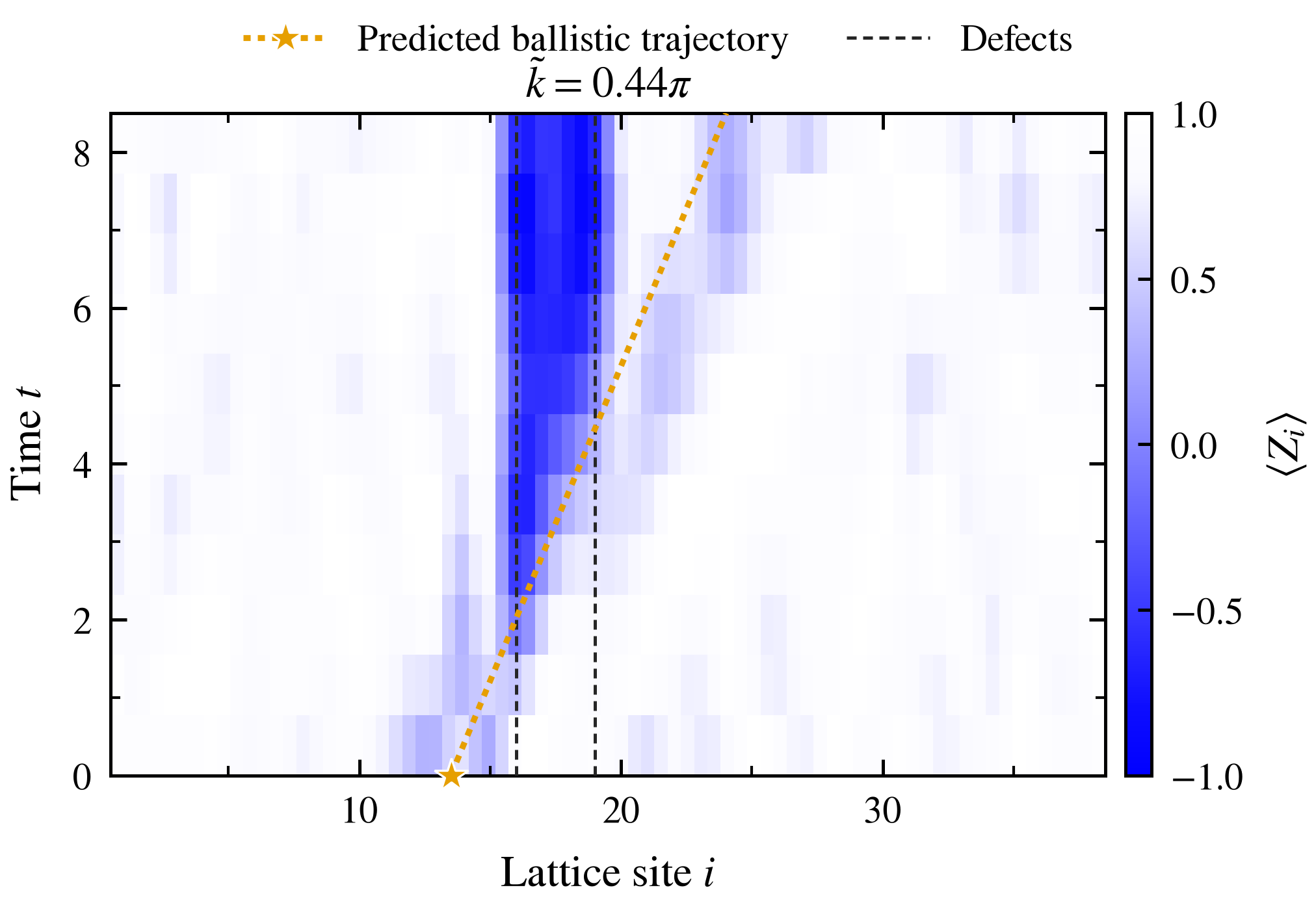}
    \caption{Hardware data for the Ising model at $\tilde k = 0.44\pi$, $g=0.7$: mean $Z$-basis outcome $\braket{Z_i}$ versus site and time. The wavepacket is perfectly transmitted through both Kramers--Wannier defects. All momentum sectors are shown in Fig.~\ref{fig:isingmodel}. The star indicates the initial location of the wave packet.}
    \label{fig:ising_single_panel}
\end{figure}

\begin{figure}
    \centering
    \includegraphics[width=\linewidth]{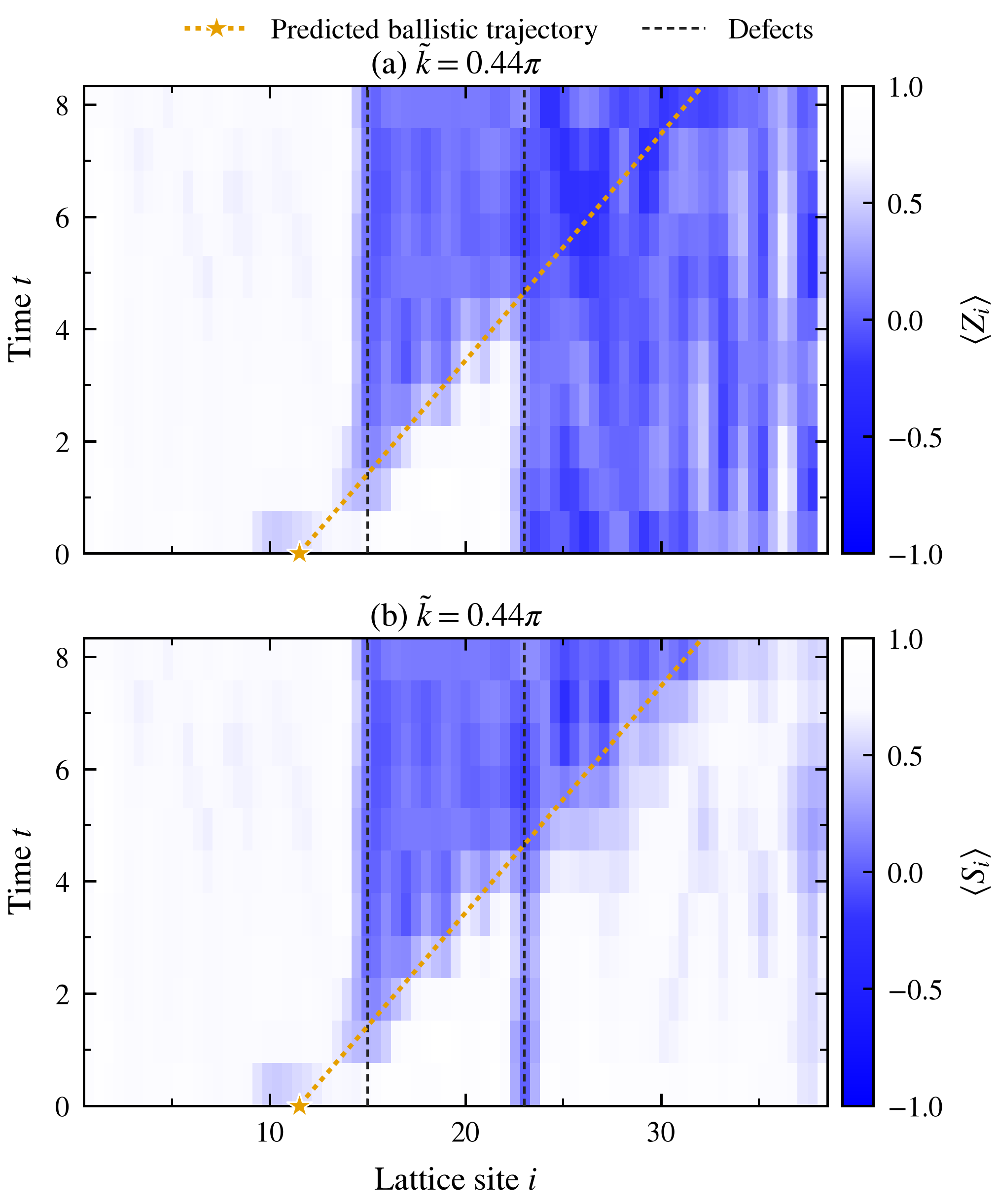}
    \caption{Hardware data for the cluster model at $\tilde k = 0.44\pi$, $g=0.7$: (a) mean $Z$-basis outcome $\braket{Z_i}$ and (b) string order $\braket{S_i}$. Past $I_R$ the wavepacket is not visible in (a) but carries on in (b). All momentum sectors are shown in Fig.~\ref{fig:clustermodel}.}
    \label{fig:cluster_single_panel}
\end{figure}

\begin{figure*}[tb]
    \centering
    \includegraphics[width=\linewidth]{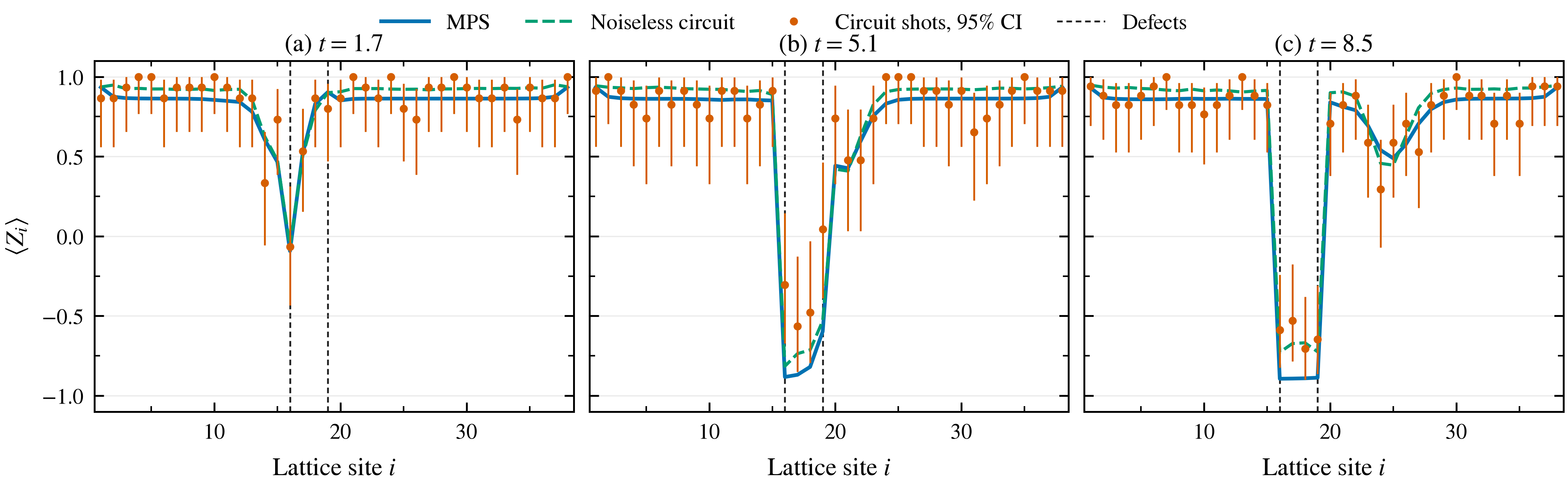}
    \includegraphics[width=\linewidth]{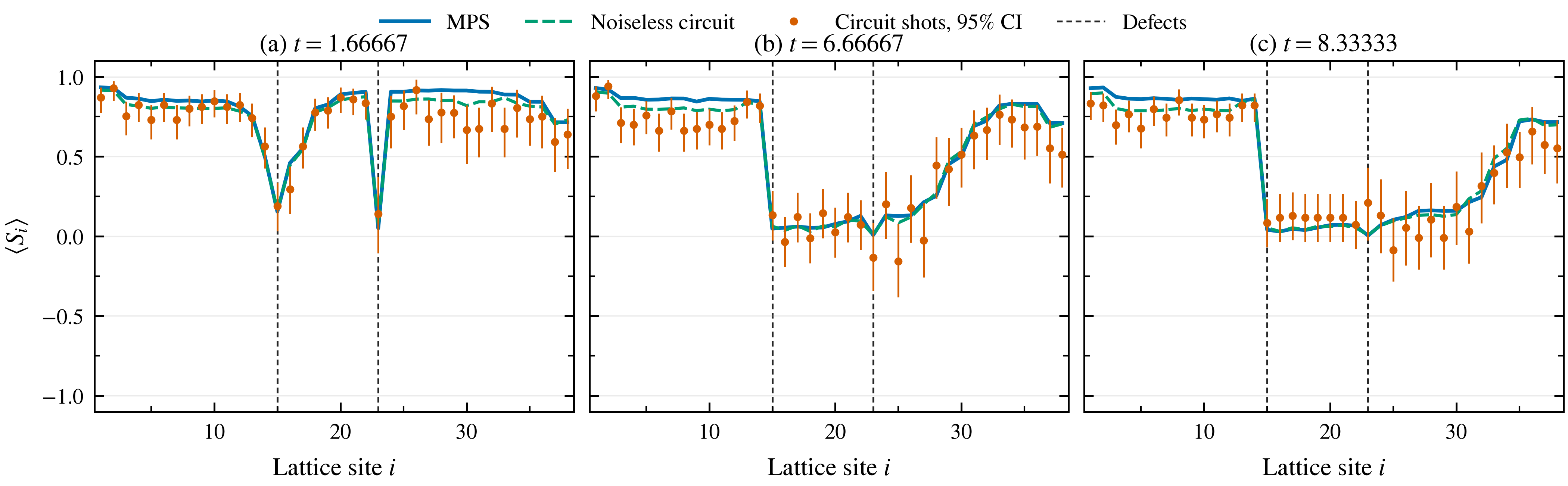}
    \caption{Time slices from Fig.~\ref{fig:ising_single_panel} and Fig.~\ref{fig:cluster_single_panel} (b) showing the comparison between the hardware shots (with Clopper–Pearson 95\% confidence intervals), noiseless simulation of the quantum circuit and a MPS TDVP simulation for $\tilde k = 0.44\pi$. We see strong agreement between the results from the quantum device and the classical simulation.}
    \label{fig:slices}
\end{figure*}
\subsection{Results}
\label{section:results}
We now test the defining prediction of duality-defect scattering: whether a wavepacket continues through the interface while changing from a local excitation into a domain-wall or string excitation. 

We perform both experiments on Quantinuum's 56-qubit H2 trapped-ion quantum computer. The Ising and cluster-model are both considered for $g=0.7$ and we compute the time evolution to a range of times using different numbers of trotter steps given by Table~\ref{tab:trotter_steps}. We choose $k_0$ to ensure that at least one of the momentum sectors lies on an optimal choice of central momentum to minimise the spread of group velocities (as discussed in Appendix~\ref{app:optimalk}). Figs. \ref{fig:ising_single_panel}
 and \ref{fig:cluster_single_panel} show the hardware results for this optimum $\tilde k$.
 
Both experiments use a wavepacket width of $M=8$ on a system with $L=36$ spin lattice sites. Post-classifying the circuit outcomes by the measured ancilla bits gives eight momentum sectors for the Ising model and four for the cluster model, where the reduction follows from the degeneracy associated with the two interleaved Ising chains (discussed in Appendix~\ref{app:optimalk}).

Figures~\ref{fig:isingmodel} and \ref{fig:clustermodel} show the resulting space-time trajectories. Reversing the sign of the post-classified momentum reverses the direction of propagation, while changing its magnitude changes the trajectory slope, as expected from the momentum-dependent quasiparticle group velocity. A Gaussian wavepacket in position space also has a finite momentum width and consequently the momentum sectors with small group velocity contain components with both signs of velocity and display broader bidirectional signals.

Figure~\ref{fig:ising_single_panel} shows the scattering of the Ising wavepacket through the KW defects. The wavepacket is initially prepared as a superposition of local Pauli-$X$ excitations in $H_\mathrm{p}$. Its diagonal space-time trajectory continues across the interface into $H_\mathrm{dw}$, where KW duality maps the excitation to a non-local domain wall. No separate reflected branch is resolved at the interface, and the continuation of the trajectory is consistent with the transmission expected for the ideal duality defect. Because the domain wall remains detectable through the local Ising order parameter, it can still be followed after entering the dual region. Crossing the second defect returns the wavepacket to a local representation. We contrast Fig.~\ref{fig:ising_single_panel} with Fig.~\ref{fig:mpsising} which demonstrates that when the model and dual model do not match across the defect we do not observe perfect transmission, and for a large enough difference in parameters we observe perfect reflection.

Figure~\ref{fig:cluster_single_panel} provides the clearest distinction between propagation and local observability. Before the wavepacket crosses the second defect, its trajectory is visible in the local $Z$-basis signal. After it enters the cluster region, the propagating feature disappears from this local observable. Taken alone, the loss of local contrast could be interpreted as absorption, decoherence, or reflection at the interface. The non-local string-order measurements show instead that the trajectory continues into the SPT region with the same direction of propagation. The particle has therefore not simply disappeared; the defect has transformed it into an excitation whose presence is encoded non-locally along a string extending from the interface. This is the lattice mechanism corresponding to the transfer of an outgoing particle into the twisted sector invoked in resolutions of the monopole paradox.

A useful feature of post-classification is that one family of circuits provides access to every momentum sector shown in Figs.~\ref{fig:isingmodel} and \ref{fig:clustermodel}. Each outcome satisfying the unary Hamming-weight constraint is retained and assigned to its corresponding value of $k$. 

Figure~\ref{fig:slices} highlights particular time slices from Fig.~\ref{fig:isingmodel} and Fig.~\ref{fig:clustermodel} for the optimum central momentum to minimise the spread of group velocities. We can see how the hardware results closely agree with the classical MPS simulation for both models.

\section{Discussion}

We have implemented the scattering of momentum-resolved wavepackets through duality defects on a programmable trapped-ion quantum computer. These defects are not conventional impurity potentials, but interfaces between different operator descriptions of the same underlying quantum theory. Their defining effect is therefore not merely a change in the velocity or amplitude of a wavepacket, but a change in the type of operator required to describe the outgoing excitation.

In the Ising experiment, a wavepacket prepared from local spin excitations propagates through a Kramer-Wannier defect and enters the dual region as a domain wall associated with the endpoint of a disorder string. The continued space-time trajectory and the absence of a separately resolved reflected branch are consistent with the perfect-transmission property of the ideal defect. Because the domain wall remains detectable through a local order parameter, this construction permits the particle-to-string-to-particle sequence across the two defects to be followed directly.

The Cluster model experiment provides a sharper distinction between propagation and local observability. Once the excitation enters the SPT region, it disappears from the local $Z$ signal but remains visible through non-local string order. The complementary behaviour of these measurements shows that the wavepacket has not simply been lost at the interface. Instead, its information has been transferred into a non-local excitation extending across multiple sites. In the language of duality defects, the same physical excitation has entered a sector in which it no longer admits the original local particle description.

This process realises, in a controllable quantum many-body system, the mechanism proposed to resolve the monopole-scattering paradox. In monopole scattering, the apparent absence of an allowed outgoing particle can be understood by recognising that the final state belongs to a twisted sector and remains non-locally connected to the monopole. Here the duality defect plays the role of the sector boundary: an incoming local excitation crosses the interface and emerges as the endpoint of a string. 

The experiment also gives an operational role to dualities and generalised symmetries. Rather than using a duality only as a formal relation between Hamiltonians, we use its defect as a dynamical device that converts experimentally accessible local states into otherwise costly non-local states. The wavepacket is prepared where it is local, and the defect generates its non-local representation through time evolution. This viewpoint may extend to other MPO dualities and to systems containing symmetry defects, disorder operators, anyonic strings, or topological boundary conditions.

Taken together, the results show a particle crossing a duality interface and changing from a local excitation into a domain-wall or string excitation. They provide a quantum-computing implementation of the particle-to-twisted-sector conversion underlying the modern resolution of the monopole paradox and demonstrate that an apparently missing particle can remain present in observables that are intrinsically non-local.

{\it{Acknowledgements}}
A.U. was supported by FWO Junior Postdoctoral Fellowship (grant No.\ 12AHS26N). FV acknowledges funding from the UKRI grant EP/Z003342/1, BOFGOA
(Grant No. BOF23/GOA/021) and IBOF (Grant No.~IBOF23/064).
Experimental
data reported in this work was produced by Quantinuum
H2-2 quantum computer, powered by Honeywell, in September 2026. Access to the device
was provided through the Quantinuum promotional access program. We are extremely grateful to the Quantinuum technical and support team for the support they
provided during this project.
\bibliography{ref}

@article{o2025quantum,
  title={Quantum state preparation via piecewise QSVT},
  author={O'Brien, Oliver and S{\"u}nderhauf, Christoph},
  journal={Quantum},
  volume={9},
  pages={1786},
  year={2025},
  publisher={Verein zur F{\"o}rderung des Open Access Publizierens in den Quantenwissenschaften}
}

@article{turco2025creation,
  title={Creation of wave packets for quantum chromodynamics on quantum computers},
  author={Turco, Matteo and Quinta, Gon{\c{c}}alo and Seixas, Jo{\~a}o and Omar, Yasser},
  journal={Physical Review D},
  volume={112},
  number={3},
  pages={034506},
  year={2025},
  publisher={APS}
}

@article{jain2024quantum,
  title={Quantum Fourier networks for solving parametric PDEs},
  author={Jain, Nishant and Landman, Jonas and Mathur, Natansh and Kerenidis, Iordanis},
  journal={Quantum Science and Technology},
  volume={9},
  number={3},
  pages={035026},
  year={2024},
  publisher={IOP Publishing}
}

@article{cooleytukey,
 ISSN = {00255718, 10886842},
 URL = {http://www.jstor.org/stable/2003354},
 author = {James W. Cooley and John W. Tukey},
 journal = {Mathematics of Computation},
 number = {90},
 pages = {297--301},
 publisher = {American Mathematical Society},
 title = {An Algorithm for the Machine Calculation of Complex Fourier Series},
 urldate = {2026-05-29},
 volume = {19},
 year = {1965}
}

@article{Callan:1982ah,
    author = "Callan, Jr., Curtis G.",
    title = "{Disappearing Dyons}",
    reportNumber = "Print-82-0006 (PRINCETON)",
    doi = "10.1103/PhysRevD.25.2141",
    journal = "Phys. Rev. D",
    volume = "25",
    pages = "2141",
    year = "1982"
}

@article{Rubakov:1982fp,
    author = "Rubakov, V. A.",
    title = "{Adler-Bell-Jackiw Anomaly and Fermion Number Breaking in the Presence of a Magnetic Monopole}",
    doi = "10.1016/0550-3213(82)90034-7",
    journal = "Nucl. Phys. B",
    volume = "203",
    pages = "311--348",
    year = "1982"
}

@article{childs2021theory,
  title={Theory of trotter error with commutator scaling},
  author={Childs, Andrew M and Su, Yuan and Tran, Minh C and Wiebe, Nathan and Zhu, Shuchen},
  journal={Physical Review X},
  volume={11},
  number={1},
  pages={011020},
  year={2021},
  publisher={APS}
}

@article{suzuki1991general,
  title={General theory of fractal path integrals with applications to many-body theories and statistical physics},
  author={Suzuki, Masuo},
  journal={Journal of mathematical physics},
  volume={32},
  number={2},
  pages={400--407},
  year={1991},
  publisher={American Institute of Physics}
}

@article{suzuki1976generalized,
  title={Generalized Trotter's formula and systematic approximants of exponential operators and inner derivations with applications to many-body problems},
  author={Suzuki, Masuo},
  journal={Communications in Mathematical Physics},
  volume={51},
  number={2},
  pages={183--190},
  year={1976},
  publisher={Springer}
}

@article{layden2022first,
  title={First-order trotter error from a second-order perspective},
  author={Layden, David},
  journal={Physical Review Letters},
  volume={128},
  number={21},
  pages={210501},
  year={2022},
  publisher={APS}
}

@article{van_beest_fermion-monopole_2024,
    title = {Fermion-monopole scattering in the Standard Model},
    volume = {2024},
    issn = {1029-8479},
    url = {https://link.springer.com/10.1007/JHEP08(2024)004},
    doi = {10.1007/JHEP08(2024)004},
    number = {8},
    urldate = {2025-09-30},
    journal = {Journal of High Energy Physics},
    author = {Van Beest, Marieke and Smith, Philip Boyle and Delmastro, Diego and Mouland, Rishi and Tong, David},
    month = aug,
    year = {2024},
    pages = {4},
}

@article{van_beest_monopoles_2025,
    title = {Monopoles, scattering, and generalized symmetries},
    volume = {2025},
    issn = {1029-8479},
    url = {https://doi.org/10.1007/JHEP03(2025)014},
    doi = {10.1007/JHEP03(2025)014},
    number = {3},
    journal = {Journal of High Energy Physics},
    author = {van Beest, Marieke and Smith, Philip Boyle and Delmastro, Diego and Komargodski, Zohar and Tong, David},
    month = mar,
    year = {2025},
    pages = {14},
}

@article{loladze_monopole-fermion_2025,
    title = {Monopole-{Fermion} {Scattering} and the {Solution} to the {Semiton}–{Unitarity} {Puzzle}},
    volume = {134},
    issn = {0031-9007, 1079-7114},
    url = {https://link.aps.org/doi/10.1103/PhysRevLett.134.051602},
    doi = {10.1103/PhysRevLett.134.051602},
    number = {5},
    urldate = {2025-12-07},
    journal = {Physical Review Letters},
    author = {Loladze, Vazha and Okui, Takemichi},
    month = feb,
    year = {2025},
    pages = {051602},
}

@article{loladze_dynamics_2025,
  title={Dynamics of the fermion-rotor system},
  author={Loladze, Vazha and Okui, Takemichi and Tong, David},
  journal={Journal of High Energy Physics},
  volume={2026},
  number={1},
  pages={52},
  year={2026},
  publisher={Springer}
}

@article{Polchinski:fermion_rotor,
    author = "Polchinski, Joseph",
    title = "{Monopole Catalysis: The Fermion Rotor System}",
    reportNumber = "HUTP-84-A005",
    doi = "10.1016/0550-3213(84)90398-5",
    journal = "Nucl. Phys. B",
    volume = "242",
    pages = "345--363",
    year = "1984"
}

@article{Ueda:2026wvc,
  title = {Emergent Andreev Reflection from a Lattice Duality Defect},
  author = {Ueda, Atsushi and Numasawa, Tokiro and De Vos, Boris and Watanabe, Masataka},
  journal = {Phys. Rev. Lett.},
  volume = {137},
  issue = {14},
  pages = {146504},
  numpages = {6},
  year = {2026},
  month = {Sep},
  publisher = {American Physical Society},
  doi = {10.1103/h1rv-bv9y},
  url = {https://link.aps.org/doi/10.1103/h1rv-bv9y}
}

@article{childs2012hamiltonian,
author = {Childs, Andrew M. and Wiebe, Nathan},
title = {Hamiltonian simulation using linear combinations of unitary operations},
year = {2012},
issue_date = {November 2012},
publisher = {Rinton Press, Incorporated},
address = {Paramus, NJ},
volume = {12},
number = {11–12},
issn = {1533-7146},
journal = {Quantum Info. Comput.},
month = nov,
pages = {901–924},
numpages = {24}
}

@inproceedings{Gily_n_2019, series={STOC ’19},
   title={Quantum singular value transformation and beyond: exponential improvements for quantum matrix arithmetics},
   url={http://dx.doi.org/10.1145/3313276.3316366},
   DOI={10.1145/3313276.3316366},
   booktitle={Proceedings of the 51st Annual ACM SIGACT Symposium on Theory of Computing},
   publisher={ACM},
   author={Gilyén, András and Su, Yuan and Low, Guang Hao and Wiebe, Nathan},
   year={2019},
   month=June, pages={193–204},
   collection={STOC ’19} }

@article{mcardle2026quantum,
  title={Quantum State Preparation without Coherent Arithmetic},
  author={McArdle, Sam and Gily{\'e}n, Andr{\'a}s and Berta, Mario},
  journal={Physical Review Letters},
  volume={136},
  number={24},
  pages={240603},
  year={2026},
  publisher={APS}
}

@article{pino2021demonstration,
  title={Demonstration of the trapped-ion quantum CCD computer architecture},
  author={Pino, Juan M and Dreiling, Jennifer M and Figgatt, Caroline and Gaebler, John P and Moses, Steven A and Allman, Michael S and Baldwin, Charles H and Foss-Feig, Michael and Hayes, David and Mayer, Karl and others},
  journal={Nature},
  volume={592},
  number={7853},
  pages={209--213},
  year={2021},
  publisher={Nature Publishing Group UK London}
}

@article{corcoles2021exploiting,
  title={Exploiting dynamic quantum circuits in a quantum algorithm with superconducting qubits},
  author={C{\'o}rcoles, Antonio D and Takita, Maika and Inoue, Ken and Lekuch, Scott and Minev, Zlatko K and Chow, Jerry M and Gambetta, Jay M},
  journal={Physical Review Letters},
  volume={127},
  number={10},
  pages={100501},
  year={2021},
  publisher={APS}
}

@article{farrell2024quantum,
  title={Quantum simulations of hadron dynamics in the Schwinger model using 112 qubits},
  author={Farrell, Roland C and Illa, Marc and Ciavarella, Anthony N and Savage, Martin J},
  journal={Physical Review D},
  volume={109},
  number={11},
  pages={114510},
  year={2024},
  publisher={APS}
}

@article{berry2015simulating,
  title={Simulating Hamiltonian dynamics with a truncated Taylor series},
  author={Berry, Dominic W and Childs, Andrew M and Cleve, Richard and Kothari, Robin and Somma, Rolando D},
  journal={Physical review letters},
  volume={114},
  number={9},
  pages={090502},
  year={2015},
  publisher={APS}
}

@article{babbush2018encoding,
  title={Encoding electronic spectra in quantum circuits with linear T complexity},
  author={Babbush, Ryan and Gidney, Craig and Berry, Dominic W and Wiebe, Nathan and McClean, Jarrod and Paler, Alexandru and Fowler, Austin and Neven, Hartmut},
  journal={Physical Review X},
  volume={8},
  number={4},
  pages={041015},
  year={2018},
  publisher={APS}
}

@article{jordan2012quantum,
  title={Quantum algorithms for quantum field theories},
  author={Jordan, Stephen P and Lee, Keith SM and Preskill, John},
  journal={Science},
  volume={336},
  number={6085},
  pages={1130--1133},
  year={2012},
  publisher={American Association for the Advancement of Science}
}

@article{jordan2011quantum,
author = {Jordan, Stephen P. and Lee, Keith S. M. and Preskill, John},
title = {Quantum computation of scattering in scalar quantum field theories},
year = {2014},
issue_date = {September 2014},
publisher = {Rinton Press, Incorporated},
address = {Paramus, NJ},
volume = {14},
number = {11–12},
issn = {1533-7146},
journal = {Quantum Info. Comput.},
month = sep,
pages = {1014–1080},
numpages = {67}
}

@article{jordan2014quantum,
  title={Quantum algorithms for fermionic quantum field theories},
  author={Jordan, Stephen P and Lee, Keith SM and Preskill, John},
  journal={arXiv preprint arXiv:1404.7115},
  year={2014}
}

@article{brennen2014multi,
  title = {Multiscale quantum simulation of quantum field theory using wavelets},
  author = {Brennen, Gavin K. and Rohde, Peter and Sanders, Barry C. and Singh, Sukhwinder},
  journal = {Phys. Rev. A},
  volume = {92},
  issue = {3},
  pages = {032315},
  numpages = {11},
  year = {2015},
  month = {Sep},
  publisher = {American Physical Society},
  doi = {10.1103/PhysRevA.92.032315},
  url = {https://link.aps.org/doi/10.1103/PhysRevA.92.032315}
}

@article{barata2021single,
  title={Single-particle digitization strategy for quantum computation of a $\phi$ 4 scalar field theory},
  author={Barata, Jo{\~a}o and Mueller, Niklas and Tarasov, Andrey and Venugopalan, Raju},
  journal={Physical Review A},
  volume={103},
  number={4},
  pages={042410},
  year={2021},
  publisher={APS}
}

@article{davoudi2024scattering,
  title={Scattering wave packets of hadrons in gauge theories: Preparation on a quantum computer},
  author={Davoudi, Zohreh and Hsieh, Chung-Chun and Kadam, Saurabh V},
  journal={Quantum},
  volume={8},
  pages={1520},
  year={2024},
  publisher={Verein zur F{\"o}rderung des Open Access Publizierens in den Quantenwissenschaften}
}

@article{zemlevskiy2025scalable,
  title={Scalable quantum simulations of scattering in scalar field theory on 120 qubits},
  author={Zemlevskiy, Nikita A},
  journal={Physical Review D},
  volume={112},
  number={3},
  pages={034502},
  year={2025},
  publisher={APS}
}

@article{ueda2025,
	author = {Ueda, Atsushi and Vander Linden, Vic and De Vos, Boris and Lootens, Laurens and Haegeman, Jutho and Fendley, Paul and Verstraete, Frank},
	date = {2026/10/01},
	doi = {10.1038/s41567-026-03390-5},
	id = {Ueda2026},
	isbn = {1745-2481},
	journal = {Nature Physics},
	number = {10},
	pages = {1684--1690},
	title = {Perfect particle transmission through duality defects},
	url = {https://doi.org/10.1038/s41567-026-03390-5},
	volume = {22},
	year = {2026}}

@article{lootens2023dualities,
  title={Dualities in one-dimensional quantum lattice models: Symmetric Hamiltonians and matrix product operator intertwiners},
  author={Lootens, Laurens and Delcamp, Clement and Ortiz, Gerardo and Verstraete, Frank},
  journal={PRX Quantum},
  volume={4},
  number={2},
  pages={020357},
  year={2023},
  publisher={APS}
}

@article{lootens2024dualities,
  title={Dualities in one-dimensional quantum lattice models: Topological sectors},
  author={Lootens, Laurens and Delcamp, Clement and Verstraete, Frank},
  journal={PRX Quantum},
  volume={5},
  number={1},
  pages={010338},
  year={2024},
  publisher={APS}
}

@article{li2023noninvertible,
  title={Noninvertible duality transformation between symmetry-protected topological and spontaneous symmetry breaking phases},
  author={Li, Linhao and Oshikawa, Masaki and Zheng, Yunqin},
  journal={Physical Review B},
  volume={108},
  number={21},
  pages={214429},
  year={2023},
  publisher={APS}
}

@article{robertson2025approximate,
author = {Robertson, Niall and Akhriev, Albert and Vala, Jiri and Zhuk, Sergiy},
title = {Approximate Quantum Compiling for Quantum Simulation: A Tensor Network Based Approach},
year = {2025},
issue_date = {September 2025},
publisher = {Association for Computing Machinery},
address = {New York, NY, USA},
volume = {6},
number = {3},
url = {https://doi.org/10.1145/3731251},
doi = {10.1145/3731251},
journal = {ACM Transactions on Quantum Computing},
month = may,
articleno = {20},
numpages = {15} 
}

@article{schollwock2011density,
  title={The density-matrix renormalization group in the age of matrix product states},
  author={Schollw{\"o}ck, Ulrich},
  journal={Annals of physics},
  volume={326},
  number={1},
  pages={96--192},
  year={2011},
  publisher={Elsevier}
}

@article{PhysRevLett.69.2863,
  title = {Density matrix formulation for quantum renormalization groups},
  author = {White, Steven R.},
  journal = {Phys. Rev. Lett.},
  volume = {69},
  issue = {19},
  pages = {2863--2866},
  numpages = {0},
  year = {1992},
  month = {Nov},
  publisher = {American Physical Society},
  doi = {10.1103/PhysRevLett.69.2863},
  url = {https://link.aps.org/doi/10.1103/PhysRevLett.69.2863}
}

@article{van2021real,
  title={Real-time scattering of interacting quasiparticles in quantum spin chains},
  author={Van Damme, Maarten and Vanderstraeten, Laurens and De Nardis, Jacopo and Haegeman, Jutho and Verstraete, Frank},
  journal={Physical Review Research},
  volume={3},
  number={1},
  pages={013078},
  year={2021},
  publisher={APS}
}

@article{lootens2025entanglement,
  title={Entanglement and the density matrix renormalization group in the generalized Landau paradigm},
  author={Lootens, Laurens and Delcamp, Clement and Verstraete, Frank},
  journal={Nature Physics},
  volume={21},
  number={10},
  pages={1657--1663},
  year={2025},
  publisher={Nature Publishing Group UK London}
}

@article{PhysRevX.13.041052,
  title = {A Race-Track Trapped-Ion Quantum Processor},
  author = {Moses, S. A. and Baldwin, C. H. and Allman, M. S. and Ancona, R. and Ascarrunz, L. and Barnes, C. and Bartolotta, J. and Bjork, B. and Blanchard, P. and Bohn, M. and Bohnet, J. G. and Brown, N. C. and Burdick, N. Q. and Burton, W. C. and Campbell, S. L. and Campora, J. P. and Carron, C. and Chambers, J. and Chan, J. W. and Chen, Y. H. and Chernoguzov, A. and Chertkov, E. and Colina, J. and Curtis, J. P. and Daniel, R. and DeCross, M. and Deen, D. and Delaney, C. and Dreiling, J. M. and Ertsgaard, C. T. and Esposito, J. and Estey, B. and Fabrikant, M. and Figgatt, C. and Foltz, C. and Foss-Feig, M. and Francois, D. and Gaebler, J. P. and Gatterman, T. M. and Gilbreth, C. N. and Giles, J. and Glynn, E. and Hall, A. and Hankin, A. M. and Hansen, A. and Hayes, D. and Higashi, B. and Hoffman, I. M. and Horning, B. and Hout, J. J. and Jacobs, R. and Johansen, J. and Jones, L. and Karcz, J. and Klein, T. and Lauria, P. and Lee, P. and Liefer, D. and Lu, S. T. and Lucchetti, D. and Lytle, C. and Malm, A. and Matheny, M. and Mathewson, B. and Mayer, K. and Miller, D. B. and Mills, M. and Neyenhuis, B. and Nugent, L. and Olson, S. and Parks, J. and Price, G. N. and Price, Z. and Pugh, M. and Ransford, A. and Reed, A. P. and Roman, C. and Rowe, M. and Ryan-Anderson, C. and Sanders, S. and Sedlacek, J. and Shevchuk, P. and Siegfried, P. and Skripka, T. and Spaun, B. and Sprenkle, R. T. and Stutz, R. P. and Swallows, M. and Tobey, R. I. and Tran, A. and Tran, T. and Vogt, E. and Volin, C. and Walker, J. and Zolot, A. M. and Pino, J. M.},
  journal = {Phys. Rev. X},
  volume = {13},
  issue = {4},
  pages = {041052},
  numpages = {25},
  year = {2023},
  month = {Dec},
  publisher = {American Physical Society},
  doi = {10.1103/PhysRevX.13.041052},
  url = {https://link.aps.org/doi/10.1103/PhysRevX.13.041052}
}

@article{samanta2025dualitytwist,
  author  = {Samanta, Sutapa and Wang, Derek S. and
             Rahmani, Armin and Mitra, Aditi},
  title   = {Isolated zero mode in a quantum computer from a duality twist},
  journal = {Quantum},
  volume  = {9},
  pages   = {1957},
  year    = {2025},
  doi     = {10.22331/q-2025-12-30-1957}
}

@article{aasen2016topological,
  author  = {Aasen, David and Mong, Roger S. K. and Fendley, Paul},
  title   = {Topological defects on the lattice: {I}. The {Ising} model},
  journal = {Journal of Physics A: Mathematical and Theoretical},
  volume  = {49},
  number  = {35},
  pages   = {354001},
  year    = {2016},
  doi     = {10.1088/1751-8113/49/35/354001}
}

@misc{shao2023noninvertible,
  author        = {Shao, Shu-Heng},
  title         = {What's done cannot be undone:
                   {TASI} lectures on non-invertible symmetries},
  year          = {2023},
  eprint        = {2308.00747},
  archivePrefix = {arXiv},
  primaryClass  = {hep-th},
  url           = {https://arxiv.org/abs/2308.00747}
}

@misc{khan2024quantumoperations,
  author        = {Khan, Maaz and Khan, Syed Anausha Bin Zakir and
                   Mohd, Arif},
  title         = {Quantum operations for {Kramers--Wannier} duality},
  year          = {2024},
  eprint        = {2405.09361},
  archivePrefix = {arXiv},
  primaryClass  = {hep-th},
  url           = {https://arxiv.org/abs/2405.09361}
}

@article{else2013hidden,
  author  = {Else, Dominic V. and Bartlett, Stephen D. and
             Doherty, Andrew C.},
  title   = {The hidden symmetry-breaking picture of
             symmetry-protected topological order},
  journal = {Physical Review B},
  volume  = {88},
  pages   = {085114},
  year    = {2013},
  doi     = {10.1103/PhysRevB.88.085114}
}

@article{farrell2026digital,
	author = {Farrell, Roland C. and Zemlevskiy, Nikita A. and Illa, Marc and Preskill, John},
	date = {2026/09/11},
	doi = {10.1038/s41567-026-03436-8},
	id = {Farrell2026},
	isbn = {1745-2481},
	journal = {Nature Physics},
	title = {Digital quantum simulations of scattering in quantum field theories using W states},
	url = {https://doi.org/10.1038/s41567-026-03436-8},
	year = {2026}}

\clearpage
\appendix
\section{Experimental implementation}
\label{app:experimental_implementation}
To implement the scattering dynamics of both models on quantum hardware, we
\begin{enumerate}
    \item prepare an approximation to the many-body ground state,
    \item excite a local wavepacket on top of this background,
    \item evolve the wavepacket through the duality defects, and
    \item measure the outgoing excitation using the appropriate local or non-local observable.
\end{enumerate}
Each step is adapted to reduce the circuit depth and measurement cost on current quantum hardware as discussed below.

\subsection{Ground state preparation}
\begin{figure}
    \centering
    \includegraphics[width=0.8\linewidth]{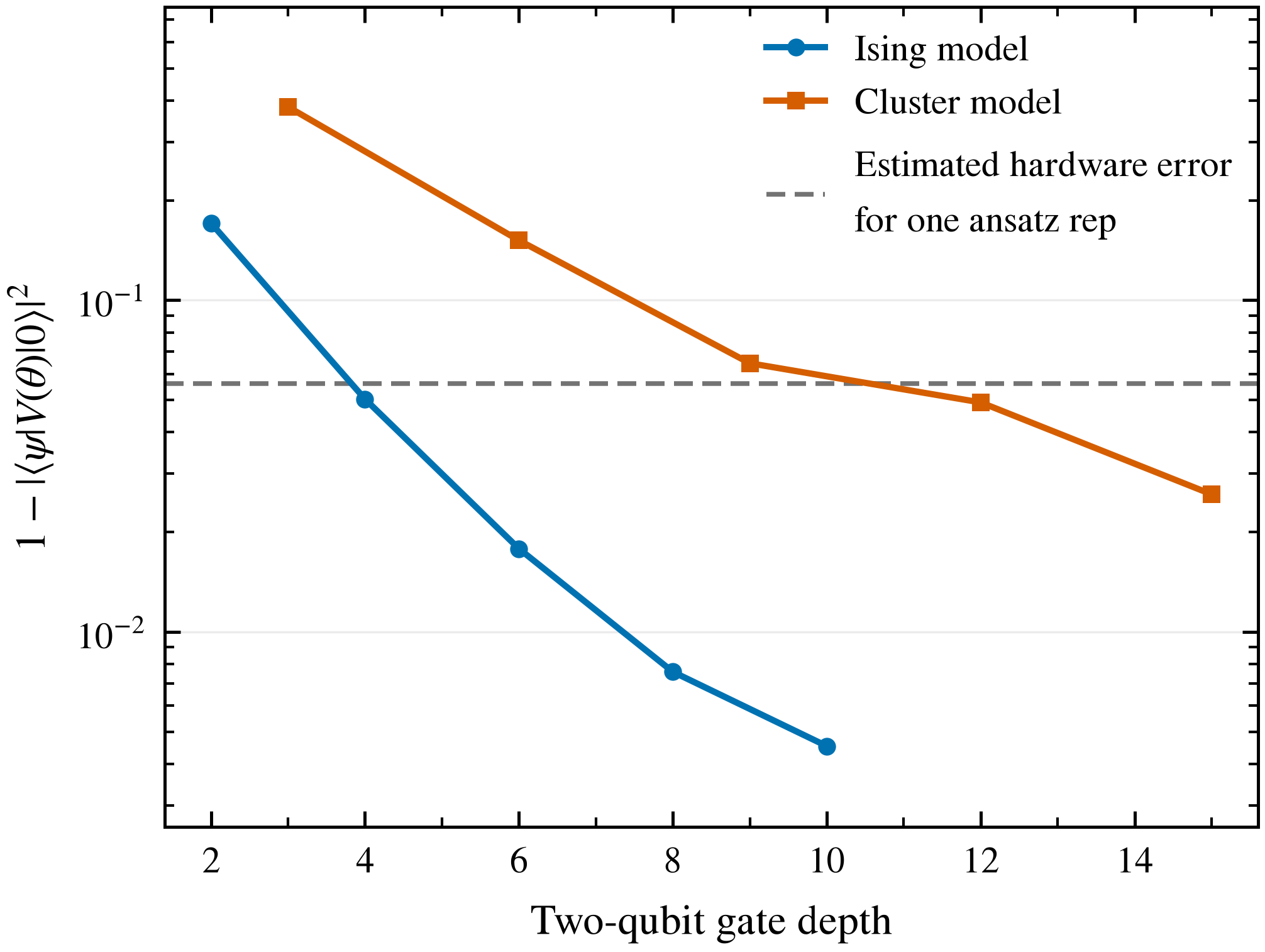}
    \caption{Ground-state preparation error for the Ising model and Cluster model on a 36-qubit system at $g=0.7$ as a function of the two-qubit depth of the optimised hardware-efficient ansatz $V(\theta)$. The vertical axis shows the infidelity between the variationally prepared state $V(\theta)\ket{0}$ and the target DMRG/MPS ground state $\ket{\psi}$. We also show the estimated error due to hardware noise introduced by including one additional layer in the ansatz.}
    \label{fig:groundstateprep}
\end{figure}

Matrix-product-state (MPS) representations of the ground states can be found efficiently and to high precision using DMRG \cite{PhysRevLett.69.2863,schollwock2011density}. We then use AQCtensor \cite{robertson2025approximate} to obtain low-depth circuits that approximate these states. The method begins with a parameterised hardware-efficient ansatz, compresses it into an MPS using tensor-network simulation, and evaluates its fidelity with the target ground-state MPS. Standard numerical optimisation, such as L-BFGS-B, then converges to a low-depth approximation. We use a low-depth ansatz circuit tailored to the connectivity of the model of interest. For the Ising model, we use pairwise rZZ interactions interspersed with rX rotations. For the Cluster model, we use a tailored ansatz of two separate chains of next-nearest neighbour rZZ interactions until $I_R$ when we transition into a single chain of pairwise rZZ interactions, again interspersed with rX rotations. Figure~\ref{fig:groundstateprep} shows the dependence of the ground-state preparation error on circuit depth. Based on this data, we use an ansatz with a two-qubit depth of 8 for the Ising model and 9 for the cluster model. 
\subsection{Wavepacket preparation}
We implement the protocol of Section~\ref{section:localized_wavepacket} with several modifications for current quantum hardware.

First, the standard implementation of \textsc{SELECT} requires expensive multiplexed controlled unitaries. We instead use a unary encoding of the ancilla register. At the cost of additional ancillas, this reduces \textsc{SELECT} to $M$ two-qubit \textsc{CNOT} gates. In the $H_\mathrm{p}$ region, where the incident wavepacket is created, the localised excitation $\phi_i^\dagger$ is a Pauli $X$ operator on site $i$ for both models.

Second, for Gaussian wavepackets, \textsc{PREP} prepares the state
\begin{equation}
    \frac{1}{\mathcal{N}}\sum_{x=0}^{M-1}e^{-\left(\frac{x-\frac{M-1}{2}}{\sigma}\right)^2}e^{\frac{2\pi i}{M}k_0 x} \ket{2^x}\,.
\end{equation}
Because the Gaussian tails decay exponentially, we can take $M\ll L$ with little truncation error. We approximate \textsc{PREP} by numerically optimising a hardware-efficient variational ansatz at modest classical cost.

The unary encoding also changes the implementation of \textsc{QFT}. We use the shallow-depth unary QFT circuit proposed in Ref.~\cite{jain2024quantum}, which replaces the radix-2 operations in the Cooley-Tukey fast Fourier transform \cite{cooleytukey} with two-qubit gates. It requires $n\log n$ two-qubit gates and benefits from the all-to-all connectivity of the trapped-ion architecture. 

Finally, the unary encoding provides a simple error-detection condition: we discard ancillary outcomes whose Hamming weight is not one.

\subsection{Time evolution}
We implement time evolution using a second-order Suzuki-Trotter decomposition \cite{suzuki1976generalized, suzuki1991general} whereby time evolution by a target Hamiltonian $H=\sum_j h_j$ is performed through the symmetrised operator:
\begin{equation}
   e^{-iHt} \approx \left(e^{\frac{ -ih_1 t}{2r}}\cdots e^{\frac{ -ih_{L-1} t}{2r}}e^{\frac{ -ih_L t}{r}} e^{\frac{ -ih_{L-1} t}{2r}}\cdots e^{\frac{ -ih_1 t}{2r}}\right)^r\,.
\end{equation}
The operator error of this approximation can be bounded as $\epsilon_T=O\left(\frac{t^3}{r^2}\sum_{j,k,l}\|[h_l,[h_j,h_k]]\|\right)$ \cite{childs2021theory}. For local models we have $\sum_{j,k,l}\|[h_l,[h_j,h_k]]\|=O(L)$ and hence evolving to time $t$ with $r$ timesteps we get $\epsilon_T=O\left(\frac{L t^3}{r^2}\right)$.

For our circuits we can choose an ordering such that our second-order trotter circuit has almost the same depth as a first-order trotter circuit \cite{layden2022first}. This is seen by grouping our Hamiltonian into two parts ($H_\alpha$, $H_\beta$) such that every term in each part commutes with every other term in the part. Then we can see that the second-order trotter circuit can be written as:
\begin{equation}
    \left(e^{\frac{ -iH_\alpha t}{2r}}e^{\frac{ -iH_\beta t}{r}} e^{\frac{ -iH_\alpha t}{2r}}\right)^r = e^{\frac{ -iH_\alpha t}{2r}}\left(e^{\frac{ -iH_\beta t}{r}} e^{\frac{ -iH_\alpha t}{r}}\right)^r e^{\frac{ iH_\alpha t}{2r}}\,.
\end{equation}
Therefore, we can achieve second-order trotter error scaling with ``almost" first-order trotter circuits. We further reduce the depth of our circuits by ensuring every trotter step is optimally parallelised.
\subsection{Measurement}
In the Ising model, the outgoing domain wall can be detected through a local order parameter, so we measure all system qubits in the $Z$ basis. In the cluster model, detecting the transmitted excitation is instead detected through the non-local order parameters $S_i$
\begin{equation}
S_i = \begin{cases}Z_{I_R}Z_{i}\prod_{j=1}^{\frac{i-I_R-1}{2}}X_{I_R+2j} & \text{for odd } (i-I_R)\\
    X_{I_R}Z_{i}\prod_{j=1}^{\frac{i-I_R-2}{2}}X_{I_R+2j+1} &\text{for even }  (i-I_R)\,.
    \end{cases}
    \label{eq:Si}
\end{equation}
Strings terminating on sites of the same parity commute qubit-wise. Therefore, splitting them into even and odd groups allows all string-order expectation values to be evaluated using two measurement settings:
\begin{equation}
    Z_{I_R}\prod_{\substack{j=I_R+1\\ j-I_R\ \text{odd}}}^{L-2} Z_{j}X_{j+1} \;\text{ and }\;  X_{I_R}\prod_{\substack{j=I_R+1\\ j-I_R\ \text{odd}}}^{L-2} X_{j}Z_{j+1}\,.
\end{equation}
The non-local wavepacket can therefore be reconstructed with two measurement circuits per evolution time.
\section{Optimal Central Momentum}
\label{app:optimalk}
For the Ising model, we choose the optimal $k_0$ to minimise the spread of the group velocities within the wavepacket. In the idealised case this would be given by $\arccos g$, however due to the localisation of our wavepacket this is no longer optimum. Using the dispersion relation for the Ising model we numerically calculated the optimum $k_0$ for our wavepacket width to be given by $\tilde k_0 = \frac{2\pi }{M}k_0=0.44\pi$.

The localised wavepacket created in the cluster model is
\begin{equation}
    \Psi_k^{\dagger} = \sum_{p=0}^{M-1} e^{-\left(\frac{p-\frac{M-1}{2}}{\sigma}\right)^2}e^{\frac{2\pi i}{M}k p} X_p \,.
\end{equation}
Because the model consists of two Ising chains on the odd and even sites, this can be written as
\begin{equation}
    \Psi_k^{\dagger} = \Psi_{k,1}^{\dagger} +\Psi_{k,2}^{\dagger}\,,
\end{equation}
where
\begin{equation}
        \Psi_{k,1}^{\dagger} = \sum_{p=0}^{\frac{M-2}{2}} e^{-\left(\frac{2p-\frac{M-1}{2}}{\sigma}\right)^2}e^{\frac{4\pi i}{M}k p} X_{p,1} 
\end{equation}
\begin{equation}
        \Psi_{k,2}^{\dagger} = \sum_{p=0}^{\frac{M-2}{2}} e^{-\left(\frac{2p+1-\frac{M-1}{2}}{\sigma}\right)^2}e^{\frac{2\pi i}{M}k (2p+1)} X_{p,2} \,.
\end{equation}
The central momentum of these wavepackets is therefore $\tilde k =\frac{4\pi}{M}k$ rather than $\frac{2\pi}{M}k$. Pairs of ancilla outcomes correspond to the same effective momentum, and we choose $\frac{2\pi }{M}k_0=0.22\pi$ to minimise the spread of group velocities.

\begin{table}[htbp]
    \centering
    \begin{tabular}{|c|c|c|c|}
    \hline
        \multicolumn{4}{|c|}{\textsc{Ising Model}} \\\hline
         \textbf{Time}& \textbf{Trotter steps}&\textbf{2 qubit gate count}&\textbf{Post-selection}  \\\hline
         0.00&0&193&96.7\%\\\hline
         1.70&3&355&75.7\%\\\hline
         3.40&6&460&78.3\%\\\hline
         5.10&9&577&88.7\%\\\hline
         6.80&12&688&86.7\%\\\hline
         8.50&15&799&91.3\%\\\hline
    \end{tabular}\vspace{1cm}
    \begin{tabular}{|c|c|c|c|}
    \hline
     \multicolumn{4}{|c|}{\textsc{Cluster Model}} \\\hline
         \textbf{Time}& \textbf{Trotter steps}&\textbf{2 qubit gate count}&\textbf{Post-selection}  \\\hline
         0.00&0&156&97.4\%\\\hline
         1.67&4&368&92.6\%\\\hline
         3.33& 7&476&90.1\%\\\hline
         5.00&10&585&91.9\%\\\hline
         6.67&14&729&91.4\%\\\hline
         8.33&17&837&90.2\%\\\hline
    \end{tabular}
    \caption{Tables illustrating the various time steps evaluated in the quantum time evolution of each model along with the number of trotter steps and 2 qubit gates used in such evaluation. The post-selection proportion corresponds to the proportion of samples that are measured with hamming weight 1 and hence are retained into the plots. This is an effective error mitigation strategy for detecting bit flip errors in the ancillas.}
    \label{tab:trotter_steps}
\end{table}

\begin{figure*}
    \centering
    \includegraphics[width=\linewidth]{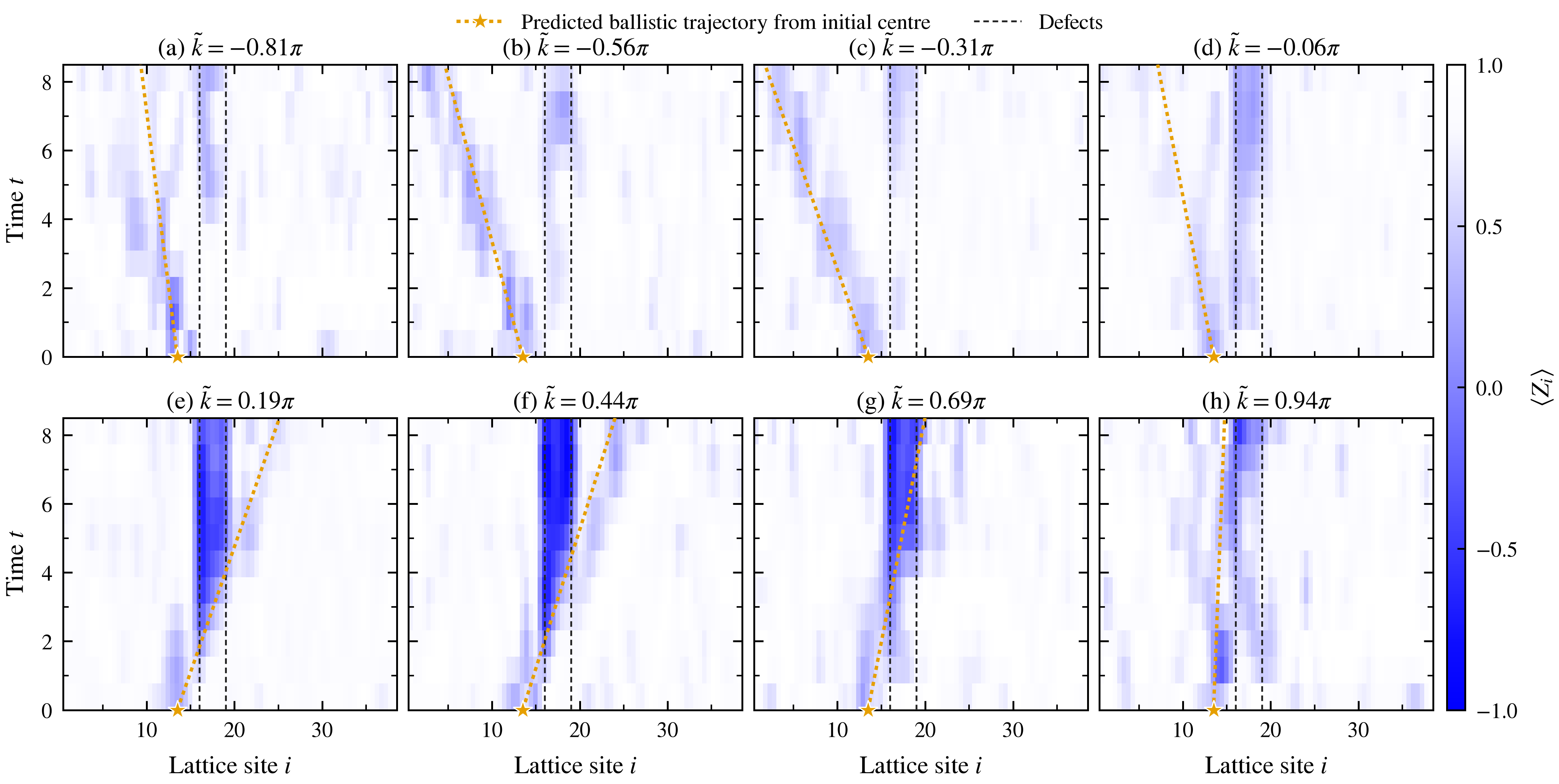}
    \caption{Momentum-resolved transmission and particle-to-domain-wall conversion across KW duality defects on the Quantinuum H2 quantum computer, with $g=0.7$, $I_L=16$, $I_R=19$, and $L=36$. Each panel shows the mean $Z$-basis measurement outcome as a function of lattice site and evolution time, conditioned on the measured unary-ancilla momentum outcome $k$ (where $\tilde k =\frac{2\pi}{M}(k+k_0)$ with $M=8$). The 300 shots performed are split between the 8 post-classified panels. The wavepacket begins as a superposition of local Pauli-$X$ excitations in $H_p$. After crossing a defect into $H_\mathrm{dw}$, it is represented as the endpoint of a non-local disorder string and is interpreted as a propagating domain wall. When it crosses back to $H_p$ it again becomes localised. The predicted ballistic trajectory is given by the group velocity of the central momentum $\tilde k$ of each wavepacket. As the wavepackets are localised in space they are spread in momentum space and hence actually contain a range of group velocities. We are using linear interpolation to smooth out the hardware plots as we only measured the six distinct timesteps detailed in Table~\ref{tab:trotter_steps}.}
    \label{fig:isingmodel}
\end{figure*}

\begin{figure*}[htb]
    \centering
    \includegraphics[width=\linewidth]{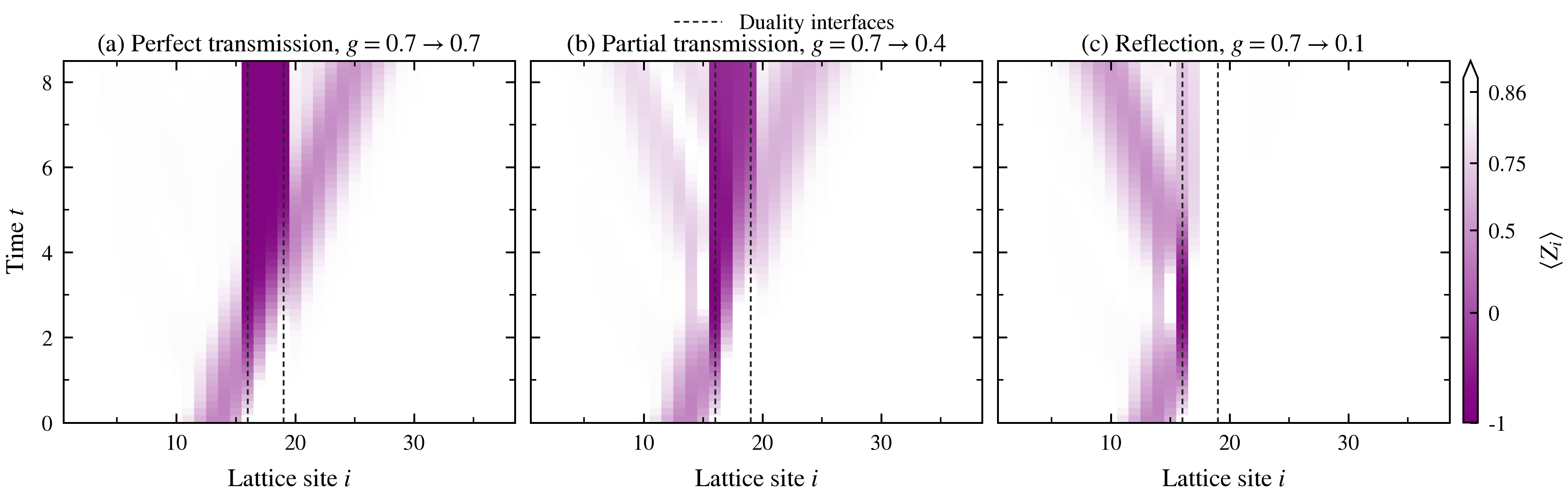}
    \caption{MPS simulation of wavepacket propagation on the Ising model with a KW duality defect for a range of $g$ values for the dual model ($H_\mathrm{dw}$). Each panel shows the interpolated mean $Z$-basis outcome as a function of lattice site and evolution time. Every panel uses the same central momentum of $\tilde k = 0.44\pi$ to minimise the spread of group velocities. Perfect transmission is observed for $g=0.7\rightarrow 0.7$ (a) in accordance with Fig.~\ref{fig:ising_single_panel}. Partial transmission is observed for $g=0.7\rightarrow 0.4$ (b) and near perfect reflection is observed for $g=0.7\rightarrow0.1$ (c).}
    \label{fig:mpsising}
\end{figure*}

\begin{figure*}
    \centering
    \includegraphics[width=\linewidth]{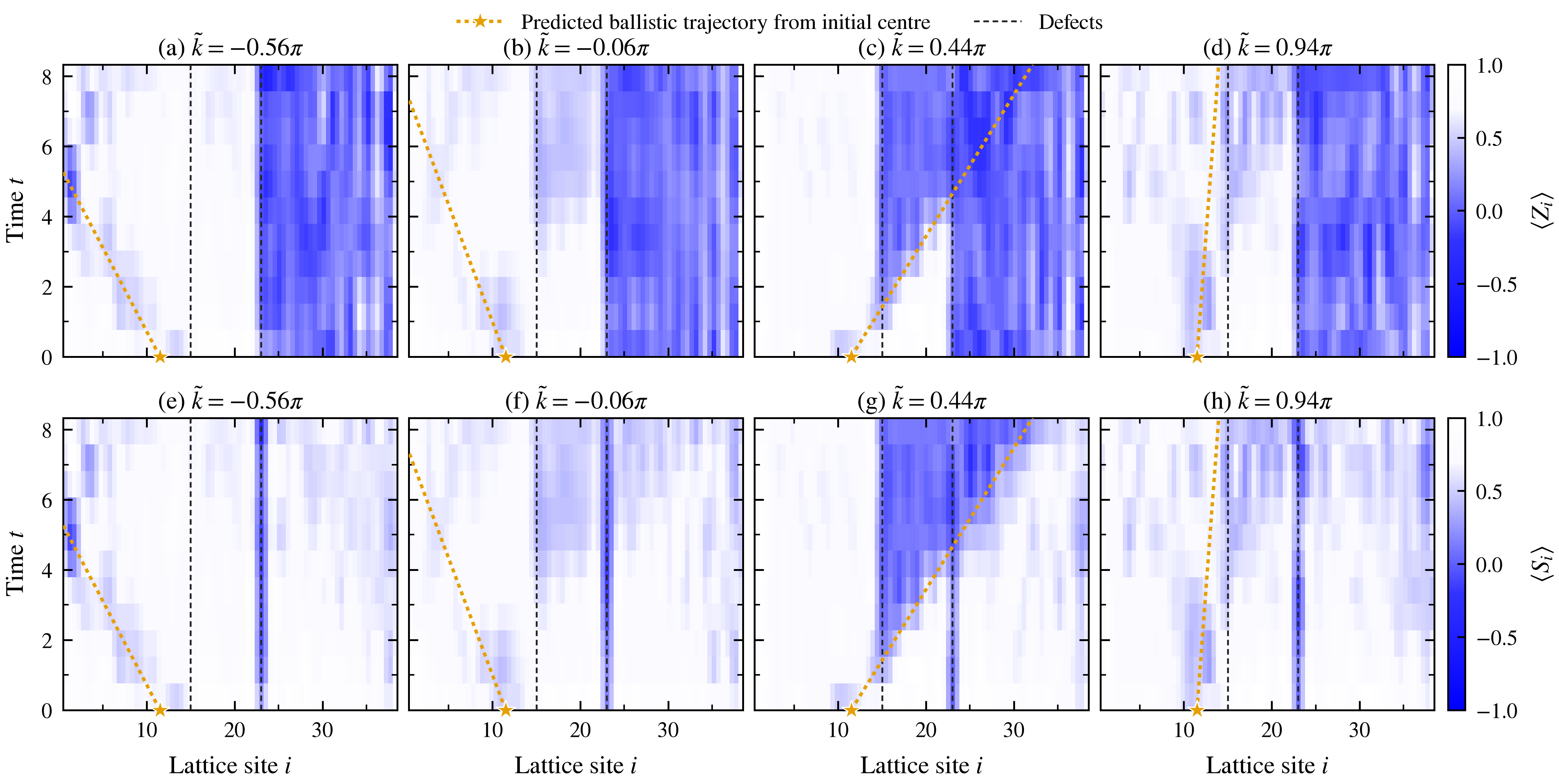}
    \caption{Conversion of a locally visible particle into a non-local string excitation across sequential duality defects of $H_c$ on the Quantinuum H2 quantum computer, with $g=0.7$, $I_L=15$, $I_R=23$, and $L=36$. The columns show four post-classified momentum sectors conditioned on the measured unary-ancilla momentum outcome $k$ (where $\tilde k =\frac{4\pi}{M}(k+k_0)$ with $M=8$). The upper row displays the mean $Z$-basis measurement outcome, while the lower row displays the mean non-local string-order signal $\braket{S_i}$ defined in Eq.~\ref{eq:Si} (with $S_i = Z_i$ to the left of $I_R$). Before the wavepacket enters the cluster region, its trajectory is visible locally. After crossing the second defect at $I_R$, the propagating feature disappears from the local signal but continues in the string-order measurements. The complementary signals show that the excitation has entered the SPT region while changing from a locally detectable particle into a non-local string excitation. We are using linear interpolation to smooth out the hardware plots as we only measured the six distinct timesteps detailed in Table~\ref{tab:trotter_steps}. Two measurement settings were used to produce these plots with 400 hardware shots post-classified over the four panels in each row. For sites to the left of $I_R$ the two measurement settings coincide so there are effectively 800 hardware shots split over each row.}
    \label{fig:clustermodel}
\end{figure*}

\end{document}